\documentclass[prl,twocolumn,showpacs,preprintnumbers,showkeys]{revtex4-1}
\usepackage{amsmath}
\usepackage{amssymb}
\usepackage{graphicx}
\usepackage{dcolumn}
\usepackage{bm}
\usepackage{natbib}
\usepackage{color}

\newcommand{\leb}{\left (}
\newcommand{\reb}{\right )}
\newcommand{\les}{\left [}
\newcommand{\res}{\right ]}
\newcommand{\ii}{\text{i}}
\newcommand{\dd}{\text{d}}
\newcommand{\ee}{\text{e}}

\newcommand{\dens}{\rho}     
\newcommand{\modz}{r}        
\newcommand{\argz}{\varphi}  
\newcommand{\modH}{b}        
\newcommand{\argH}{\psi}     

\newcommand{\of}[1]{\leb {#1} \reb}
\newcommand{\soq}{\text{soq}}
\newcommand{\chim}{\text{ch}}
\newcommand{\strength}{\varepsilon}
\newcommand{\length}{2\ell}
\newcommand{\hlength}{\ell}

\begin{document}

\title{Self-Emerging and Turbulent Chimeras in Oscillator Chains}

\author{Grigory Bordyugov}\email{Grigory.Bordyugov@uni-potsdam.de}
\author{Arkady Pikovsky}
\author{Michael Rosenblum}
\affiliation{Institut f\"ur Physik und Astronomie,
             Universit\"at Potsdam,
             Karl-Liebknecht-Stra\ss e 24/25,
             14476 Potsdam, Germany}
\date{\today}

\begin{abstract}
We report on a self-emerging chimera state in a homogeneous chain of nonlocally and nonlinearly coupled oscillators.
This chimera, i.e., a state with coexisting regions of complete and partial synchrony, emerges via a supercritical bifurcation from a homogeneous state and thus does not require preparation of special initial conditions.
We develop a theory of chimera based on the equations for the local complex order parameter in the Ott-Antonsen approximation.
Applying a numerical linear stability analysis, we also describe the instability of the chimera and transition to phase turbulence with persistent patches of synchrony.

\end{abstract}


\newcommand{\mycolorred}{}
\newcommand{\mycolorblue}{}
\newcommand{\mycolorblack}{}


\maketitle
Populations of coupled oscillators is a paradigmatic model of nonlinear science, with numerous applications from purely physical ones like Josephson junction arrays and coupled lasers to biologically 
and even socially important~\cite{Kuramoto,*PikRosKur,*Sync,*KuramotoReview}. 
While in globally coupled ensembles and in networks one is mostly interested in the features of synchronization and desynchronization, spatially extended oscillating systems demonstrate a variety of pattern-forming phenomena.
One of the most spectacular recent findings are the so-called {\it chimera} states (CSs) which are observed in otherwise completely synchronizable oscillatory media if the system starts from specially prepared initial state.
CSs are characterized by the coexistence of regions with locally synchronized oscillators and regions where the oscillators phases are not locked but yet not completely incoherent.
CSs were initially discovered and explained theoretically in~\cite{KuramotoChimera}, and then received more analytical treatment in~\cite{StrogatzChimera,*StrogatzChimeraIJBC}.
Following those pioneering works on CSs, a large body of observations and analysis of similar regimes has been recently published, see~\cite{ShiKu04,*Ka07,*Sa06,*OmeMaTa08,*SeSaAt08,*AbMiStroWe08,*La09,*La09PhysD} and references therein.

In this Letter, we add another species to the zoo of chimeras.
The crucial difference is that our CS does not require special initial conditions: 
It emerges from a general initial state and is thus denoted as self-emerging.
CS is stable close to the bifurcation, but with a further variation 
of the parameter it becomes turbulent, so that synchronous and partially synchronous patches intermingle irregularly.
The key elements of our model are Stuart-Landau oscillators, coupled through an exponentially decaying kernel as in the original chimera setup, but with a difference that the coupling is {\em nonlinear} in the sense of~\cite{SOQ,*PikRosPhysD}.
First, we numerically demonstrate the existence of CS and then explain it in the phase dynamics framework with the help of reduced equations for the local order parameter.
Our main theoretical tools are the equations for the complex order parameter in the so-called Ott-Antonsen (OA) approximation~\cite{OttAntonsen,*OttAntonsenStab}. 
It exploits a parametrization of the probability density for ensembles of sinusoidally coupled phase oscillators and results in a closed equation for the order parameter.
The OA ansatz is closely related to the Watanabe-Strogatz theory~\cite{WSPRL,*WS} which is exact but does not yield closed equations \mycolorred in terms of the order parameter\mycolorblack.  
A connection between these two theories has been established in~\cite{PikRosPRL2008,*PikRosPart}.

As a basic model we consider a one-dimensional, periodic in space chain of the length $L~=~\length$ of nonlocally coupled Stuart-Landau oscillators
\begin{equation}
  \label{eq:slo}
  \partial_t A = \leb 1 + \ii \tilde\omega \reb A - |A|^2 A + \strength Z \;,
\end{equation}
where $A = A \of {x,t}$ is the complex amplitude, $\tilde\omega$ is the natural frequency of the oscillators,
$Z = Z\of{x,t}$ is the coupling force acting on the oscillator at $x$, and $\strength$ is a small coupling constant. The coupling is organized via a convolution of
$A\of{x,t}$ with the weight function $G\of{x} = c \ee^{-|x|}$:
\begin{equation}
  \label{eq:slof}
  B\of{x,t} = \int_{-\hlength}^\hlength \dd x' \, G\of{x-x'} A\of{x',t}\;,
\end{equation}
where the constant $c$ ensures that $\int_{-\hlength}^\hlength G\of{x}\dd x=1$. 
The forcing is then defined as 
\begin{equation}
\label{eq-for}
Z=\ee^{\ii \beta\of{|B|}} B.
\end{equation} 
\mycolorred
The phase shift $\beta \of{|B|} = \beta_0 + \beta_1 |B|^2$, chosen in spirit of~\cite{SOQ}, accounts for possible nonlinearity effects in the coupling,  i.e., the forcing $Z$ in Eq.~(\ref{eq:slo}) depends on the higher-order powers of $B$.
The weight function $G\of{x}$ is kept fixed, so that by variation of $L$ we change the ratio between the coupling width and the system size.
\mycolorblack

\begin{figure}[tb]
 \includegraphics[width=\linewidth]{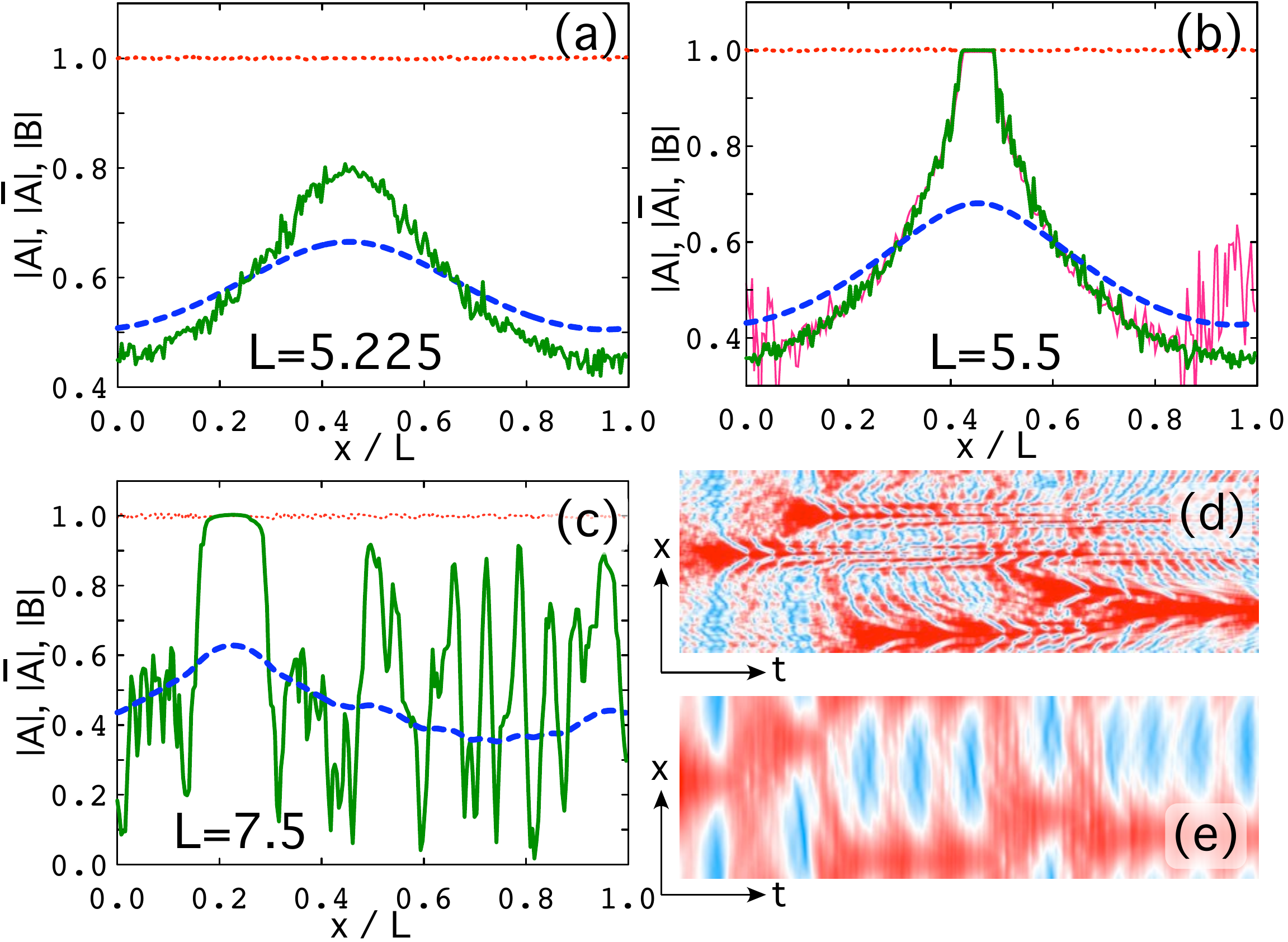}
  \caption{(color online)
  Results of simulation of Eq.~(\ref{eq:slo}). 
  (a) and (b) show stationary patterns, (c-e) show a turbulent one. 
  In (a-c) we show profiles of $|A|$ (dotted red line), $|\overline{A}|$ (solid green line), and $|B|$ (long-dashed blue line).
  The system size $L$ is specified in each frame.
  In (b), thin pink solid line (largely occulted by the green one) shows a snapshot from simulation with specially prepared initial state, see text.
  In panels  (d) and (e), showing space-time plots of $|\overline{A}|$ and $|B|$, respectively, red color (darker in b/w print) encodes values close to 1, blueish color (brighter in b/w print) encodes values close to 0.
In (d) and (e), the time span of simulations is $5000$ dimensionless time units and $L = 7.5$.
  }
  \label{fig:fig01}
\end{figure}

We integrated Eqs.~(\ref{eq:slo},\ref{eq:slof}) using $2^{17}$ sites in $x$ with $\strength = 0.01$,
$\tilde\omega = 0$, $\beta_0 = 0.4\pi$ and $\beta_1 = \leb \pi/2 - \beta_0\reb / 0.36$ (this choice of $\beta_1$ will be apparent later on) using the Runge-Kutta 4th order scheme.
Initial conditions were chosen close to the completely desynchronized state.
To measure the synchronization between neighboring oscillators on a mesoscopic scale, we calculated the coarse-grained $\overline{A}\of{x,t}$ by averaging $A\of{x,t}$ over $2^{10}$ \mycolorred closest \mycolorblack neighbors.
We varied the system size $L$ and looked at the dynamics of $|B|$ and $|\overline{A}|$.
For small $L$, we observed a spatially homogeneous, uniformly rotating, and partially synchronous state with $|\overline{A}| = |B| \approx 0.6$, like in globally coupled ensembles~\cite{SOQ}.
This state becomes unstable if the system size $L$ exceedes the critical value $L_c \approx 5.1$.
For $L > L_c$, a spatially modulated profile of $|B|$ emerged, see Fig.~\ref{fig:fig01} (a).
The profile of $|\overline{A}|$ was also stationary up to finite-size fluctuations.
Close to the transition, synchronization was only partial, with $|\overline{A}| < 1$ for all $x$.
However, for $L \gtrsim 5.35$, the profile of $|\overline{A}|$ reached unity (see Fig.~\ref{fig:fig01} (b)), which means that locally all oscillators in that region are completely synchronized, whereas in the regions with $|\overline{A}| < 1$ the local synchronization was only partial. Such state is a stationary chimera~\cite{KuramotoChimera,StrogatzChimera}.
With further increase of $L$, this regime becomes unstable and evolves into a turbulent state where synchronized patches with $|\overline{A}|\approx 1$ appeared at random places and disappeared after some lifetime, see Fig.~\ref{fig:fig01} (c)-(e). 
We call this state a \textit{turbulent chimera}. 
Below, we present a theoretical description of stationary CSs in terms of a reduced phase model.

%
%

For a small coupling constant $\strength$ we can assume that the amplitude of each Stuart-Landau oscillator remains unperturbed and the system can be described solely by its phase $\theta(x,t)$.
In the thermodynamic limit, we describe the population with a probability density distribution $\dens = \dens \of{x,\theta,t}$.
This density obeys the continuity equation $\partial_t \dens + \partial_\theta \leb \dens v \reb = 0$ with the velocity field 
$v \of{x,\theta, t}=\dot{\theta}=  \omega + \text{Im} \leb Z \of{x,t} \ee^{-\ii \theta} \reb,$ where $Z \of{x,t}$ is defined in Eq.~(\ref{eq-for}). (We rescaled the coupling parameter $\strength \rightarrow 1$ by rescaling time and frequency $t\to \strength t$, $\omega=\tilde\omega/\strength$). Next, we define the complex order parameter by
$$
z \of{x,t} = re^{i\varphi}= \int_0^{2\pi} \dd \theta \, \ee^{\ii \theta} \dens \of {x,\theta, t},
$$
so that $B \of{x,t}$ is expressed as
\begin{equation}
  \label{eq:Z}
  B \leb x, t \reb=\modH\of{x,t} \ee^{\ii \argH\of{x,t}} = \int_{-\hlength}^\hlength \dd x' \, G \leb |x - x'| \reb z \leb x', t \reb.
\end{equation}
Now we apply the theories developed in~\cite{WS,OttAntonsen,PikRosPRL2008} and use the 
OA equation for the complex order parameter  $z \of {x,t}$ 
\begin{equation}
  \label{eq:main}
  \partial_t z = \ii \omega z + \frac{1}{2} \leb Z - z^2 Z^* \reb.
\end{equation}
Together with Eqs.~(\ref{eq-for},\ref{eq:Z}) it constitutes a closed system which we analyze to study pattern formation in the domains of various length $L$.
\mycolorred
This system is similar to Eqs.~(13,14) in \cite{La09PhysD} with the exception of the nonlinear phase shift $\beta$.
\mycolorblack

\mycolorblue
The applicability of the OA theory requires justification, since for identical oscillators under a common forcing the ensemble dynamics can go beyond the OA theory~\cite{PikRosPart}.
For non-identical oscillators, the phase distribution function asymptotically fulfills the conditions for the OA theory (one says that a solution is attracted to the invariant OA manifold) \cite{OttAntonsen}.
Here, even though the oscillators are identical, the force that acts on them is inhomogeneous in space (see profiles of $|B|$ in Fig.~\ref{fig:fig01}), which plays the same mixing role as the non-identity of oscillators in~\cite{OttAntonsen} and the phase distribution will tend to the OA manifold as well.
To support this claim numerically, we ran an additional simulation with initial condition that strongly violated the OA ansatz.
This simulation also resulted in a stationary CS.
The profile of $|B|$ was indistinguishable from the previous simulation with nearly uniform initial distribution of phases.
The profile of $|\overline{A}|$ (thin solid pink line in Fig.~\ref{fig:fig01} (b)) is close to the previous one up to the finite-size fluctuations in the region of small $|B|$ values.
The differences in $|\overline{A}|$ are mostly pronounced in the domain with nearly uniform forcing $|B|$, what agrees with the aforementioned reasoning.
We also followed the deviation from the OA manifold looking at the quantity $|\langle \ee^{\ii\theta}\rangle^2 - \langle \ee^{\ii 2 \theta} \rangle|$ ($\langle \cdot \rangle$ denote averaging over small $x$-neighborhood as described above), which vanishes on the OA manifold.
We found that even for initial conditions off the OA manifold, asymptotically it always descreased in time.
\mycolorblack

With a proper choice of $\beta_0$ and $\beta_1$, Eq.~(\ref{eq:main}) supports three types of homogeneous steady states:
(i) The completely desynchronized one $z_0 = 0$, (ii) the fully synchronized one $z_1 = \ee^{\ii \Omega_1 t}$ with $\Omega_1 = \omega + \sin \of{\beta_0 + \beta_1}$, and (iii) the intermediate regime of self-organized quasiperiodicity (SOQ)~\cite{SOQ} $z_\soq = r_\soq \ee^{\ii \Omega_\soq t}$ with $r_\soq = \sqrt{\frac{\pi/2 - \beta_0}{\beta_1}}$ and $\Omega_\soq = \omega + \frac{1}{2}\leb 1 + r_\soq^2 \reb$.
Due to the normalization of the kernel $G$, $z_0$, $z_1$, and $z_\soq$ persist for all $\hlength$.
Knowing from~\cite{SOQ} that both $z_0$ and $z_1$ are unstable, we focus on $z_\soq$.
A linear stability analysis of $z_\soq$ with the perturbation $\propto y \ee^{\lambda t+\ii kx}$ results in the eigenvalue problem
$$
\lambda
\begin{pmatrix} y \ee^{\ii kx} \\ y^* \ee^{-\ii kx} \end{pmatrix}= 
  \frac{1}{2}
  \begin{pmatrix}
    \mathcal{L}_{11} & \mathcal{L}_{12}\\
    \mathcal{L}^*_{12} & \mathcal{L}^*_{11}
  \end{pmatrix}
\begin{pmatrix} y \ee^{\ii kx} \\ y^* \ee^{-\ii kx} \end{pmatrix},
$$
where
$\mathcal{L}_{11} = \leb \ii \leb r_\soq^2-1 \reb + \alpha\of{k, \hlength} \leb \ii - \beta_1 r_\soq^2 \leb 1 - r_\soq^2 \reb \reb\reb$ and $\mathcal{L}_{12} = \alpha\of{k, \hlength} \leb \ii r_\soq^2 - \beta_1 \leb 1 - r_\soq^2 \reb \reb$
with the $\hlength$-dependent wavenumber factor
\begin{equation}
  \label{eq:factor}
\alpha\of{k, \hlength} = \frac{1 + \ee^{-\hlength} \leb k \sin\of{k\hlength} - \cos\of{k\hlength}\reb}{\leb 1 - \ee^{-\hlength}\reb\leb 1 + k^2 \reb} \;,
\end{equation}
which converges to $\leb 1 + k^2 \reb^{-1}$ as $\hlength \rightarrow \infty$. 
The eigenvalues $\lambda_{1,2}$ of matrix $\mathcal{L}$ depend on both $k$ and $\hlength$.
On the infinite domain $\hlength = \infty$, we find that $\text{Re}\,\lambda_{1}\of{k, \infty} > 0$ and $\text{Im} \, \lambda_{1,2}\of{k, \infty} = 0$ for $0 < k < k_c$ with the critical wavenumber $k_c = \sqrt{\frac{\pi - 2 \beta_0}{\beta_1 + \beta_0 - \pi/2}}$.
With a finite $\hlength$, the critical wavenumber can differ from $k_c$ since $\hlength$ enters $\alpha\of{k, \hlength}$.
The critical system length $L_c = 2\hlength_c$ at which the instability occurs, is determined by the condition $\text{Re}\, \lambda\of{\pi/\hlength_c, \hlength_c} = 0$.
\begin{figure}[tb]
  \includegraphics[width=0.7\linewidth]{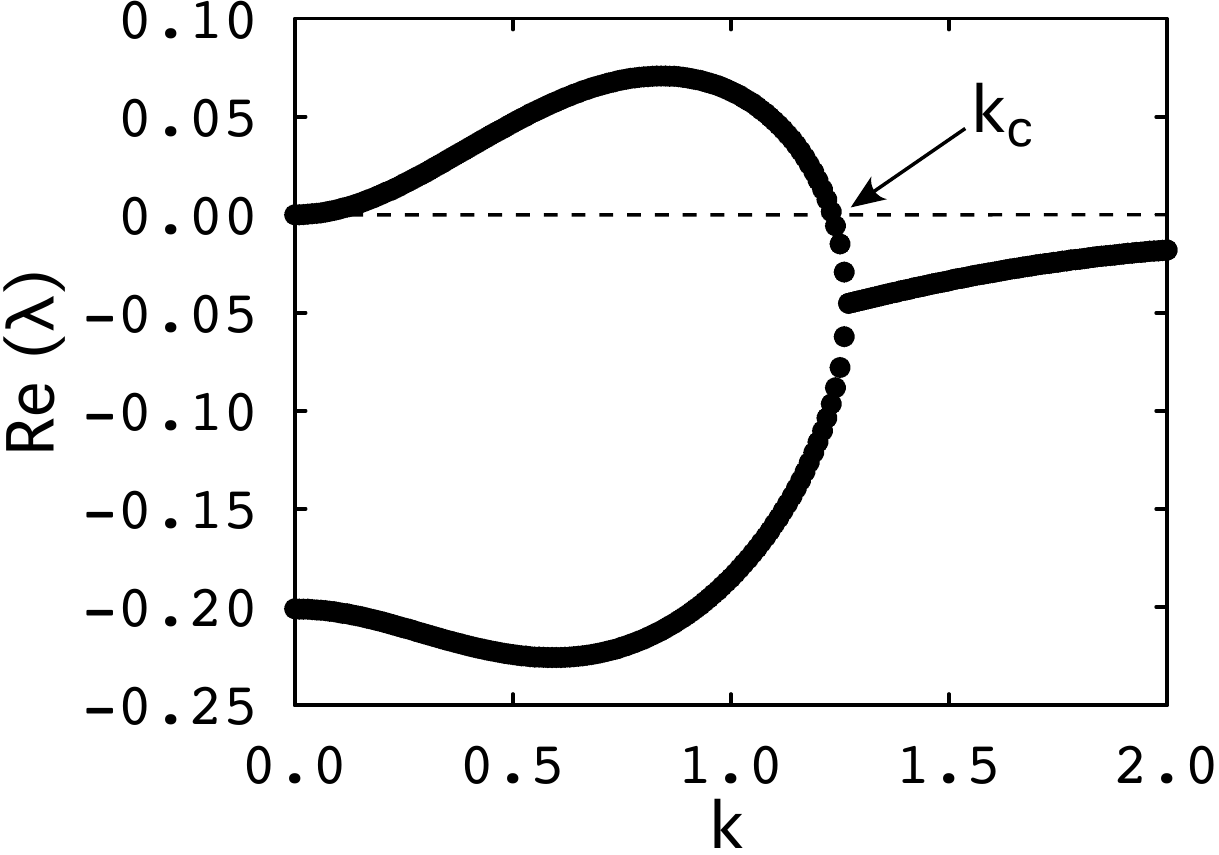}
  \caption{
  The real part of the stability eigenvalues $\lambda$ of $z_\soq$ in dependence on $k$ (the wavenumber of perturbation). 
  $k_c \approx 2\pi / 5.10$ denotes the critical wavenumber.
  }
  \label{fig:fig02}
\end{figure}
To exemplify the instability, we chose $\beta_0 = 0.4\pi$ and $\beta_1$ such that $\beta_0 + \beta_1 \times 0.6^2 = \pi/2$, so that $r_\soq = 0.6$ (cf. numerics of Fig.~\ref{fig:fig01}).
Solving the eigenvalue problem, we obtained the critical length $L_c \approx 5.09$ in a nice correspondence with the results of the direct numerical simulations.

We expect that the instability described above results in uniformly rotating spatially inhomogeneous regimes, thus we look for a solution $z_\chim \of{x,t} = \modz\of{x}\ee^{\ii \les \Omega t +\argz\of{x}\res}$ with some unknown frequency $\Omega$ and spatial functions $\modz\of{x}$ and $\argz\of{x}$.
For the forcing we assume the ansatz $B\of{x,t} = \modH\of{x} \ee^{\ii \les \Omega t + \argH\of{x} \reb}.$
When substituting this in Eq.~(\ref{eq:main}), we have to differentiate two cases: (i) for regions where $\modz < 1$ the stationarity/stability conditions yield
\begin{equation}
  \begin{split}
    \modz\of{x} &= \Gamma\of{x}- \sqrt{\Gamma^2\of{x} - 1},\\
    \argz\of{x} &= \argH\of{x} + \beta \leb \modH\of{x} \reb - \pi /2\;,
\end{split}
  \label{eq:rho_m_1}
\end{equation}
with the detuning $\Delta=\Omega - \omega$; (ii) for regions with $\modz = 1$ we have
\begin{equation}
  \argz\of{x} = \argH\of{x} + \beta \leb \modH\of{x} \reb + \arcsin\Gamma\of{x},
  \label{eq:rho_e_1}
\end{equation}
where in both Eqs.~(\ref{eq:rho_m_1},\ref{eq:rho_e_1}) $\Gamma\of{x} = \Delta / b\of{x}$.
Equation~(\ref{eq:rho_m_1}) or (\ref{eq:rho_e_1}) plus Eq.~(\ref{eq:Z}) constitute a closed system, which we could not solve analytically.
Instead, we found the stationary CSs in the frame rotating with frequency $\Omega$, employing the Newton's method for Eq.~(\ref{eq:main}) discretized in $x$.
\mycolorred
In the Newton's calculations, an additional phase pinning condition was imposed in order to pick up a unique solution from the family of phase rotations. 
The number of unknowns and equations was then balanced by taking $\Omega$ as an additional unknown.
\mycolorblack
After the solution $z_\chim\of{x}$ had been found with a desired numerical accuracy, we looked at its linear stability by computing numerically the eigenvalues $\lambda$ of the Jacobian matrix evaluated at $z_\chim\of{x}$.
Eigenvalues with positive real parts would signalize an instability of $z_\chim\of{x}$.

\begin{figure}[tb]
  \includegraphics[width=\linewidth]{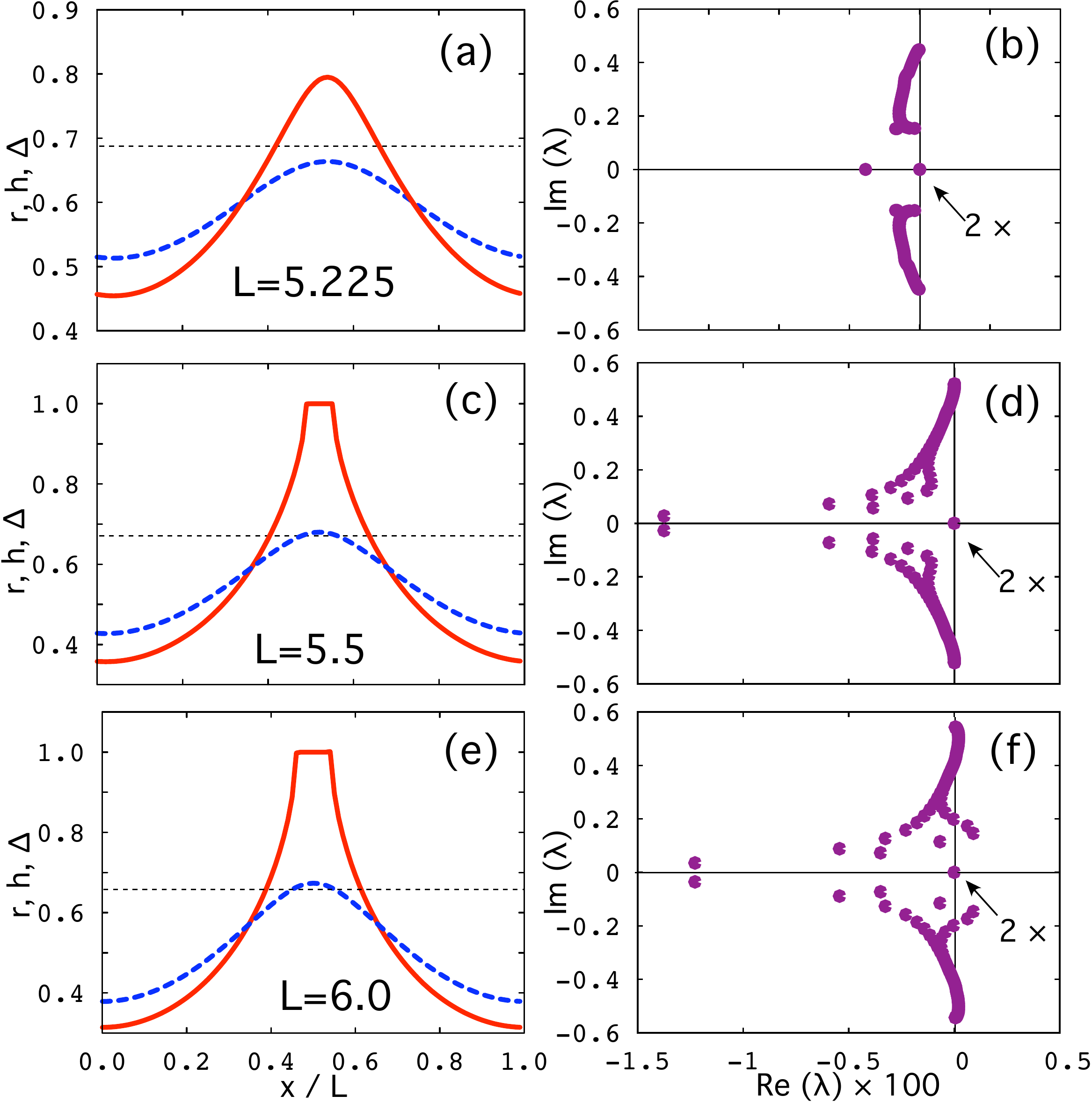}
  \caption{(color online)
  Stationary profiles of $\modz$ for different system sizes $L$.
  On the left, $\modz = |z|$ (solid red line) and $\modH$ (dashed blue line) in comparison to $\Delta$ (horizontal short-dashed line) are depicted, 
  the corresponding stability eigenvalues of the Jacobian matrix are on the right.
  }
  \label{fig:fig03}
\end{figure}

For smaller $L$ close to $L_c$, the profile of $z_\chim\of{x}$ shows a moderate sine-like modulation, see Fig.~\ref{fig:fig03} (a).
For larger $L$, the profile of $z_\chim\of{x}$ touches unity: A patch of complete synchronization emerges; this is a genuine CS. 
This occurs in the region with $\modH > \Delta$, see Fig.~\ref{fig:fig03} (c).
For our choice of $\beta_0$ and $\beta_1$, the regimes with $\modz < 1$ for all $x$ are linearly stable (see Fig.~\ref{fig:fig03} (a) and (b)), whereas the CS with a synchronized patch with $\modz=1$ becomes unstable as $L$ increases, cf. Fig.~\ref{fig:fig03} (c-f).
\mycolorred
We believe that this instability is inherited from the instability of $z_1$, due to the presence a plateau of complete synchronization with $|z| = 1$. 
\mycolorblack

\begin{figure}[tb]
  \begin{center}
    \includegraphics[width=\linewidth]{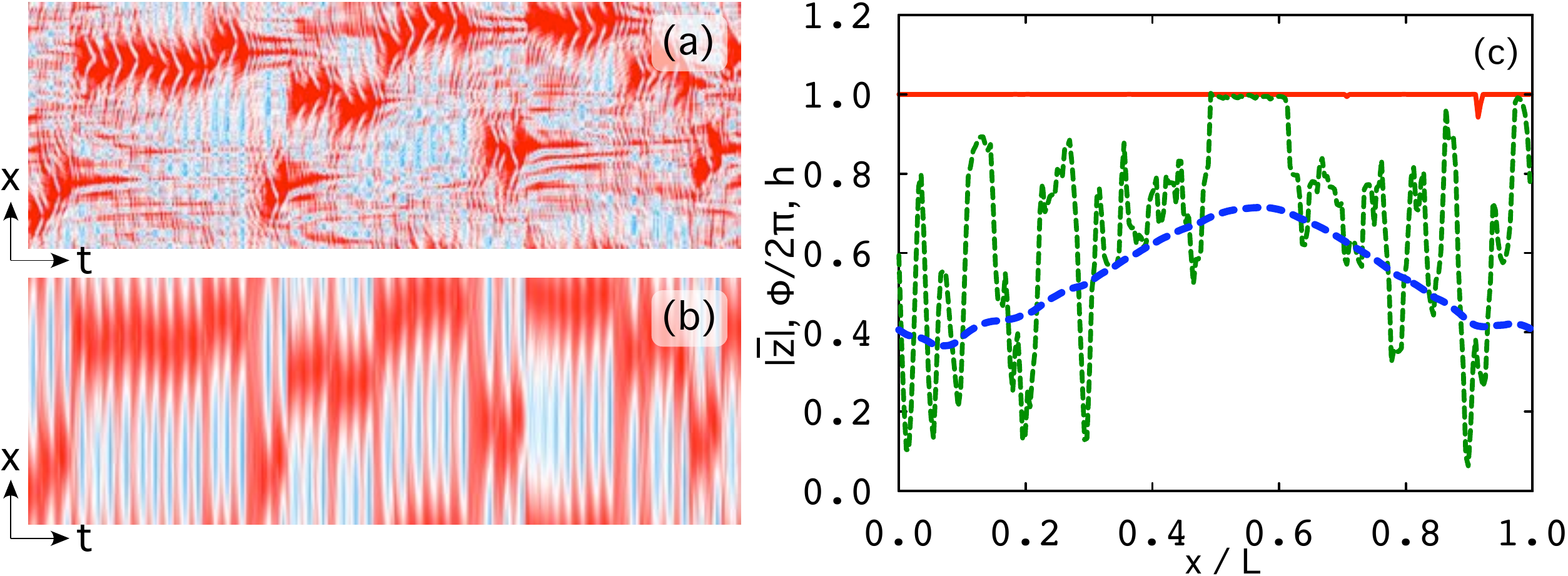}
  \end{center}
  \caption{(color online)
  Space-time plots of the coarse-grained order parameter $|\overline{z}|$ (a) 
  and of $\modH$ (b) obtained by numerical integration of Eq.~(\ref{eq:main}) with $L=7.5$.
  Red color (darker in b/w print) represents larger values close to one, bluish (brighter in b/w print) represents smaller values close to zero.
  We show simulation results from $1500$ to $3000$ dimensionless time units.
  The discretization in space $x$ was done on $1024$ sites.
  (c)
  Final snapshot of integration from (a) and (b): Solid red line shows non-coarse-grained $\modz = |z|$, green short-dashed line shows the magnitude of the coarse-grained $\overline{z}$, and long-dashed blue line represents profile of $\modH$.
  }
  \label{fig:fig04}
\end{figure}

Our numerical simulations show that  beyond the instability of the stationary CS, a turbulent state occurs, where a synchronous patch is persistent but appears at different places. 
We call this state a turbulent chimera.
Figures~\ref{fig:fig04} (a) and (b) show a result of numerical integration of our system for $L=7.5$.
Since the order parameter $z$ should be a coarse-grained quantity, in numerical simulations one has to average the numerically obtained field   over a small spatial interval (in our simulation over 16 neighboring sites) to get $\overline{z}$.
In Fig.~\ref{fig:fig04} (a) the red patches with coarse-grained $|\overline{z}|$ close to one show a larger degree of local coherence of oscillators in comparison to the blueish rest.
These coherent regions persist, allowing us to characterize the irregular state as a chimera turbulence.
The forcing magnitude $\modH$ (cf. Fig.~\ref{fig:fig03} (b)) does not show any saturation, irregularly oscillating between smaller (blue color) and larger (red color) values.
A typical final snapshot of the system is shown in Fig.~\ref{fig:fig03} (c).
There is a fully synchronous patch near $x/L\approx 0.5$ where $|\overline{z}|\approx 1$. 

The presented results can be extended to non-identical oscillators with a Lorentzian distribution of frequencies.
In this case the OA ansatz yields Eq.~(\ref{eq:main}) with an additional damping term
(so that the eigenvalue spectra of CS move from the imaginary axis into the left complex half-plane), which results in very similar states with the difference that synchronization is never complete.

Summarizing, we have demonstrated the existence of chimera-like solutions in chains of nonlocally and nonlinearly coupled oscillators.
In our system, chimera appears as a result of a long-wave instability via a supercritical bifurcation.
Unlike the previously known CSs, we do not need to prepare initial conditions carefully to avoid complete synchrony, because it is unstable.
We demonstrated that CSs are asymptotically well described within the OA theory even if the initial condition is not on the OA manifold.
As the stationary chimera becomes unstable, it evolves into a turbulent one.
In this novel regime the complex order parameter changes irregularly in space and time, but nevertheless the patches of synchronized oscillators appear persistently.
This characterizes the chimera as an important pattern in the dynamics of nonlinearly coupled oscillators.

  We acknowledge financial support from DFG via SFB 555 and useful discussions with Yu. Maistrenko, 
  E. Martens, O. Omel'chenko, M. Wolfrum, and B. Fiedler.

\end{document}